\documentclass[journal]{IEEEtran}
\usepackage{enumerate,graphicx,amsmath}

\newcommand {\mbf} [1]{\mathbf{#1}}

\ifCLASSINFOpdf
   \usepackage[pdftex]{graphicx}
\else
\fi

\begin{document}
%
\title{Key Generation: Foundations and a New  \\ Quantum Approach}
%
%
%

\author{Horace P.~Yuen
\thanks{Horace P.~Yuen is with the Department
of Electrical Engineering and Computer Science and the Department of Physics and Astronomy, Northwestern University, Evanston, IL, 60208
USA e-mail: yuen@eecs.northwestern.edu}
}

\markboth{Journal of Selected Topics in Quantum Electronics}%
{Shell \MakeLowercase{\textit{et al.}}: Bare Demo of IEEEtran.cls for Journals}
%


\IEEEspecialpapernotice{(Invited Paper)}

\maketitle

\begin{abstract}
The fundamental security and efficiency considerations for fresh key generation will be described. It is shown that the attacker's optimal probability of finding the generated key is an indispensable measure of security and that this probability limits the possibility of privacy amplification and the amount of fresh key that can be generated. A new approach to quantum cryptography to be called KCQ, keyed communication in quantum noise, is developed on the basis of quantum detection and communication theory for classical information transmission.  KCQ key generation schemes with coherent states of considerable energy
will be described. The possibility of fresh key generation is demonstrated for binary and $N$-ary detection systems under heterodyne attacks.  The security issues of these schemes will be discussed and compared with BB84. The emphasis throughout is on concrete finite bit-length protocols.
\end{abstract}


%
\IEEEpeerreviewmaketitle

\section{Introduction}
\IEEEPARstart{T}{his} paper studies the possible generation of a fresh key between two users via the process of advantage creation, which is derived from the different ciphertexts or signal observations by the user and an attacker. The quantum key distribution (QKD) protocol of BB84 and its variants \cite{bb84,gisin02,note1} are the most well-known examples, although classical scenarios of key generation were available before \cite{wyner75,ck78}. There are various problems in utilizing BB84 type protocols in concrete realistic applications, most of which can be traced to the small microscopic signals involved and the need to carry out estimation of the intrusion level for such protocols.

This paper proposes a new approach to QKD via the optimal quantum receiver principle for advantage creation: the structure of a quantum receiver that delivers the optimal performance depends on knowledge of the signal set \cite{ykl75,helstrom76}. We call this new approach \cite{yuen03} KCQ (keyed communication in quantum noise) key generation due to the explicit use of a secret key in the generation process. This KCQ approach does not exist in a classical world in which a single universal observation is optimal for all signal sets. The crucial point of KCQ in contrast to BB84 type QKD protocols is that \emph{intrusion level estimation}  may be omitted as a consequence of the optimal quantum receiver principle, which makes possible among other advantages the use of strong signals. It is hoped that KCQ would facilitate the adoption of physical key generation methods in practical optical systems. Note that KCQ key generation is in principle totally distinct from the quantum noise randomized direct encryption protocol AlphaEta ($\alpha\eta$) or Y-00 \cite{yuen03,corndorf05,hirota05}, which is KCQ direct encryption, although their implementations are closely connected.

A fresh key is, by definition, statistically independent of other system information the attacker may possess -- it has information theoretic  security according to conventional terminology \cite{note2}. However, except in the limits of none or all information on a bit sequence, it has never been made clear what operational or empirical meaning and significance the usual quantitative measures of  information theoretic security have in the context of cryptography. This is clearly an extremely serious foundational issue and it occurs in all key generation protocols, classical or quantum. The problem is compounded by the use of a shared secret key during the process of key generation which is necessary both in KCQ and in BB84 type QKD protocols where a ``public authentic channel'' has to be created. This paper will exhibit the inadequacies of the usual entropy or any single-number measure for appropriate security guarantee. {The security issues will be elaborated for realistic finite protocols.}

The case of KCQ qubit key generation will first be presented due to its close similarity to BB84, which illustrates the issues involved in a familiar context. The foundational issues of key generation in general will be discussed, especially those on proper measure of security as well as the meaning and possibility of fresh key generation via a shared secret key. The principles of KCQ key generation follow.  The coherent-state $\alpha\eta$ key generation scheme is presented next as a different way of utilizing the $\alpha\eta$ direct encryption scheme. A generalized scheme called CPPM will then be described and shown to have many desirable characteristics of a key generation protocol under the universal heterodyne attack. Effects of loss and other aspects of security will be discussed among a comparison with BB84. Some concluding remarks will be given. {The situation of theoretical security in concrete BB84 type protocols are discussed in the appendices. We focus throughout on issues important in the operation of realistic cryptosystems of any bit length. Note that the new schemes presented are far from being fully developed and merely constitute the basis of further development of this new approach of KCQ key generation.}

This paper deals with subtle issues of many facets which are often neither purely mathematical nor intuitive physical, but of a different conceptual kind \cite{yuen97,yuen05jap} that arises from the nature of cryptology, especially when the crypto mechanism involves physical principles in addition to purely mathematical ones. As in some theoretical papers in cryptology, this paper tends to be wordy and demands careful reading. It is hoped that the careful formulation described in this paper would facilitate further discussions and developments of this subtle, complicated subject of physical key generation.

\section{KCQ QUBIT KEY GENERATION}
Consider two users A ({Alice}) and B ({Bob}) and an attacker E (Eve) in a standard 4-state single-photon BB84 cryptosystem, in which each data bit is represented by one of two possible bases of
a qubit, say the vertical and horizontal states $|1\rangle$,
$|3\rangle$ and the diagonal states $|2\rangle$, $|4\rangle$. In
standard BB84, the choice of basis is revealed after {Bob} makes
his measurements, and the mismatched ones are discarded. It has
been suggested \cite{hwang98} that some advantages obtain when a
secret key is used for basis determination with usual intrusion level estimation and the resulting protocol is also secure against joint attacks
\cite{hwang03}. Clearly, no key can be generated after subtracting
the basis determination secret key if a fresh key is used
for each qubit. It was proposed in refs. \cite{hwang98,hwang03}
that a long $m$-bit secret key is to be used in a longer $n$-qubit
sequence with repetition. However, even if such use does not
affect the average information that Eve may obtain, it gives rise
to such an unfavorable distribution that security is seriously
compromised. This is because with a probability $1/2$, Eve can
guess correctly the basis of a whole block of $n/m$ qubits by
selecting the qubits where the same secret bit is used
repetitively. For a numerical illustration, let $n=10^3$ and
$m=10^2$. Then with a probability $2^{-15} \sim 0.3\times10^{-4}$, Eve can
successfully launch an opaque (intercept/resend) attack that gives
full information at the dangerous $15\%$ level \cite{gisin02} on the total bit sequence while yielding
no error to the users. In general, the strong correlation from
such repetitive use would seriously affect the appropriate
quantitative security level, and the effect of such guessing attacks on some portion of the data bit-sequence has not been accounted for with or without privacy amplification included.

This problem is alleviated when a seed key $K$ is first passed
through a pseudo-random number generator (PRNG) to yield a running key $K_r$ that is used for basis
determination, as indicated in Fig.~1. In practice, any standard
cipher running in the stream-cipher mode \cite{menezes97} can be
used as a PRNG. Even a LFSR (linear feedback shift register) is
good in the present situation. A LFSR with openly known (minimal)
connection polynomial and initial state $K$ generates a
``pseudo-random'' output with period $2^{|K|}$ \cite{menezes97}.
When a LFSR is used as a (classical) stream cipher, it is insecure
against known-plaintext attack \cite{menezes97}, in which Eve would
obtain the seed key from the running key which is itself obtained
from the input data and the output bits. However, there is no such
attack in key generation where {Alice} picks his data bits randomly.
In an attack where Eve guesses the key before measurement, the system is undermined
completely with a probability of $2^{-|K|}$. Since it is
practically easy to have $|K|\sim10^3$ or larger in a stream
cipher, such a guessing attack would have a much lower probability of success compared to, say, the
guessing attack Eve may launch by guessing the message
authentication key used to create the public channel needed in
BB84.
In contrast to the case without a PRNG, no subset of the data is vulnerable to a guessing attack that would correctly obtain a
subset of the key with a high probability.

\begin{figure} [htbp]
\begin{center}
\rotatebox{-90} {
\includegraphics[scale=0.4]{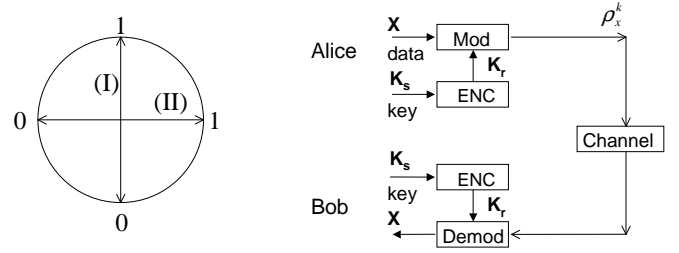}}
\caption{The qb-KCQ scheme. Left -- Two bases, I and II. Right -- Overall
encryption involves modulation with bases determined by a running key $K_r$ generated
from a seed key $K$ via an encryption mechanism denoted by the box ENC.}
\end{center}
\end{figure}

Next, we show that the seed key has complete information-theoretic
security against ciphertext-only attacks. We use \emph{upper case} for random variables and \emph{lower case} for the specific values they take. Let $\rho_x^k$ be the
quantum state corresponding to data sequence $x=x_1 \cdots x_n$
and running key $k_1 \cdots k_n$. For attacking the seed key $K$
or running key $K_r$, the quantum ciphertext reduces to $\rho^k =
\sum_x {p_x \rho^k_x}$ where $p_x$ is the apriori probability of
$x$. In our KCQ approach, we grant a full copy of the quantum
state to Eve for the purpose of bounding her information
\cite{note3}. By an optimal measurement on the qubits, Eve's
probability of correctly identifying $K$ may be obtained via $\rho^k$. Since each qubit is modulated
by its own corresponding data bit, we have $\rho_x^k =
\rho^{k_1}_{x_1} \otimes \cdots \otimes \rho_{x_n}^{k_n}$. For
uniform data commonly assumed for key generation, the $X_i$ are
independent, identically distributed (i.i.d.) Bernoulli random
variables with equal probabilities. Thus each $\rho^{k_i}=
{I_i}/2$ after averaging over $x_i$ for any value of $k_i$, with
resulting $\rho^k = \bigotimes_{i=1}^n {I_i}/2$ completely
independent of $k$. So Eve can obtain no information on $K_r$ or
$K$ at all even if she possesses a full copy of the quantum
signal. We summarize: \\ \\
\textbf{Lemma 1:} The key $K$ is completely hidden from attack in the qubit key generation scheme of Fig.~1.
\\\\
Next, we quantify the minimum security level against collective
attacks on the random data, for which Eve is assumed to have a
full copy of the quantum signal. By ``\emph{collective attack}''
we mean the situation where Eve performs a constant qubit-by-qubit
measurement on her (fictitious) full copy in the absence of any
knowledge on $K$ or $K_r$, but may employ collective classical
processing of the measurement results to take into account
correlations induced by $K$. This is analogous but different
from the usual ``collective attacks'' in BB84, because there is no
question of probe setting in the present case. Since the term
``individual attack'' in BB84 does not include collective
classical processing, our use of the term ``collective attack'' is
appropriate and allows the further generalization to joint
measurements in the most general case of ``joint attacks''
\cite{note4}
. Whatever the terminology,  $K$ or $K_r$ is actually never revealed to Eve so
that all her knowledge of the {data} must come from her
quantum measurements. Practically, so long as Eve does not have
long-term quantum memory, she would need to measure the qubits
even if she could obtain $K_r$ at a future time.

Nevertheless, solely for the purpose of lower bounding Eve's
information which is difficult to estimate otherwise because of
the correlations introduced by $K$ among the qubits, here we
conceptually grant Eve the actual $K$, and hence $K_r$, after
she made her measurements. Our KCQ principle of key generation via
optimal quantum receiver performance with versus without knowledge
of $K_r$ is easily seen to work here: Even with a full copy of the
quantum state Eve is bound to make errors in contrast to the
users. Indeed, her optimal measurement can be found by
parametrizing an arbitrary orthogonal basis which she measures,  and
optimizing the parameters assuming that $K$ is {later} granted to
her {before she makes the bit decision}. It is readily shown that general POVM measurements reduce
to orthogonal ones in this optimum binary quantum decision problem on
a qubit. Not surprisingly, her optimal error rate is $\sim
0.15$ and is obtained via the ``Breidbart'' basis \cite{bennett92}
well-known in BB84, for which one basis vector bisects the angle
between $|1\rangle$ and $|2\rangle$ or $|2\rangle$ and $|3\rangle$
depending on the bit assignment, and also in this case by the
basis obtained by rotating the Breidbart basis by $\pi/4$.

A key verification phase is to be added in a complete protocol
after error correction and privacy amplification, as discussed in Section IV B. It
does \emph{not} matter what Eve did in her interference during the
protocol execution as long as the generated key $G$ is verified. Since her information is bounded
with a full copy of the quantum state already granted to her,
there is no need for {intrusion level estimation} to ascertain her information as a function of her disturbance.

The above scheme may be generalized in many obvious ways. One is
to allow $M$ possible bases on the qubit Bloch sphere. This would
increase security without compromising efficiency as in the BB84
case, because there is no mismatched qubits to throw away and
there is no need to communicate openly what bases were measured.
It is readily shown that in the limit $M \rightarrow \infty$,
Eve's error rate goes to the maximum value 1/2 for collective
attacks \cite{yuen03}. Also, the
scheme evidently works in the \emph{same} way for Ekert type protocols
that involve shared entangled pairs. Furthermore, the same
principle may be employed for coherent-state systems with
considerable number of photons \cite{yuen03,yuen05}, as discussed in Sections V-VI.

In the present approach, error correction may be carried out by a
forward error correcting code and the resulting performance
analysis is not burdened by the need to consider Eve's probe and
whether she may hold it with quantum memory. If the channel is
estimated to have an error rate $p_c$ below 15\%, advantage is created
against collective attacks as shown above, and the existence of a
protocol that yields a net key generation rate may be carried out
asymptotically in the usual way. This channel error rate
estimation is not for advantage creation because the KCQ
principle already guarantees the users' advantage over Eve. It is
for correcting the users' channel noise and can be carried out at \emph{any}
time in contrast to {intrusion level estimation}.
Such a channel characterization is always needed in any communication line.

In particular, as long as  $p_c$ is below
the threshold $\sim 0.15$, the users could employ an error
correcting  code with rate $R$ such that
\begin{equation}
1-h_2(p_c) > R > 1-h_2(0.15),
\end{equation}
where $h_2(\cdot)$ is the binary entropy function and $1-h_2(\cdot)$ is the capacity of the corresponding (BSC) channel. The second
inequality in (1) ensures that Eve could not get at the data because the code rate $R$ exceeds her capacity. Under the first inequality
in (1) or a tighter one for concrete codes, the users can correct the channel errors {and generate fresh key at a linear rate under collective attacks as described in Section III.E.}

For concrete protocols there is
the general problem of assuring that the side information Eve has on
the error correction and privacy amplification procedures would
not allow her to obtain too much information on the generated key $G$. Under the
(unrealistic) assumption that only individual classical processing
of each qubit measurement result is made, which is the i.i.d.
assumption underlying many BB84 security analyses, Eve's Renyi
entropy, Shannon entropy, and error rate are simply related.
Quantitative results can then be easily stated as usual. For
collective and general attacks there is the problem of estimating
the Renyi entropy for applying the privacy amplification theorem
\cite{bennett95}. With {intrusion level estimation} in concrete BB84 protocols this
Renyi entropy estimate has never been carried out, {while in Renner's approach \cite{renner08,scarani08} other entropies are bounded in an unconditional security analysis to be discussed elsewhere}. The problem is
much alleviated for KCQ protocols for which a single
quantum copy is already granted to Eve for quantitatively bounding
her information {with no need of intrusion level estimation}. In particular, all the side information from error correction is accounted for by the second inequality of (1).

The complete key generation protocol, to be called qb-KCQ, is
given schematically as follows:

\begin{enumerate}[(i)]
\item {Alice} sends a sequence of $n$ random bits by a sequence of
$n$ qubit product states, each chosen randomly among two
orthogonal bases via a running key $K_r$ generated by using a PRNG
on a seed key $K$ shared by {Bob}.
\item An error control and privacy amplification procedure is employed by
the users to correct their channel errors and obtain a final
generated key $G$, while assuring , e.g.~under (1), that even after all the
associated side information and a full copy of the quantum state
is granted to Eve, errors remain for her so she has little
information on $G$.
\item The users employ key verification as in message
authentication to verify that they share the same $G$.
\end{enumerate}

The above protocol can be easily modified for performing direct
data encryption. Instead of randomly chosen bits, {Alice} sends the
data out as in (i) with error control coding but no privacy
amplification. The key verification (iii) becomes just the usual
message authentication. Note that this approach is \emph{not} possible with BB84 or its secret-key modification in refs \cite{hwang98,hwang03}, the former because of
key sifting, the latter because of the serious security breach of
Eve getting correctly a whole block of $n/m$ data bits with
probability $1/2$ described above.

In sum, the {specific features} of qb-KCQ {not obtained in} the corresponding
single-photon BB84 key generation are:

\begin{enumerate}[(1)]
\item Efficiency is increased in that there are no wasted qubits and no
need for public communication except for key verification, while
security is increased, especially for large
number of possible bases.

\item No intrusion level estimation  is required, thus no
false-alarm problem or any statistical fluctuation problem
associated with such estimation.

\item The security/efficiency analysis is unaffected even for a
multi-photon source whose output state is diagonal in the
photon-number representation, as a full copy of the
single-photon state is already granted to Eve for bounding her
information.

\item The security/efficiency quantification is similarly extended
to realistic lossy situations, while new analysis not yet
performed is otherwise needed to take into account, e.g., attacks
based on approximate probabilistic cloning \cite{fiurasek04}.

\item The security/efficiency analysis is also similarly extended to
include any side information for a finite-$n$ protocol, with no
question of holding onto the probes.

\item There are practical advantages in reducing the number of random data bits needed by {Alice} and photon counters needed by {Bob} in an
experimental implementation.

\item Direct encryption without going through key generation
first may be employed, which is impossible for BB84.

\item Sensitivity to device imperfections is reduced in the large $M$ case.
\end{enumerate}

{On the other hand,} security analysis {of this scheme has not been extensively studied as in BB84.} However, the security issues of all key generation schemes are subtle as discussed in detail in the following sections III-IV.

\section{Fundamentals of Key Generation}
This section describes the basic principles underlying all key generation schemes, classical as well as quantum. The condition of fresh key generation will be first described using the conventional entropy or mutual information criterion. The acute problem of finding operationally meaningful quantitative security criteria is then discussed in detail. This problem has not been previously treated in the literature except briefly in ref \cite{yuen06qcmc} but it affects quantum and classical protocols alike. Indeed the problem is so severe that whenever a shared secret key is needed for the key generation protocol, it is {\emph{not clear}} \emph{in what {meaningful security} sense} a fresh key has been generated. On the other hand, KCQ, BB84 and its variants, as well as classical protocols with public discussion all rely on shared secret key. This issue will be elaborated in section IV after we describe in this section how a concrete key generation scheme works in general under the criterion of Eve's optimal success probability of finding the generated key.

\subsection{Conditions for Fresh Key Generation}
A classical or quantum protocol that generates a key with information theoretic security would consist of three logical steps:

\begin{enumerate}
\item \emph{Advantage Creation}

The users A and B create a communication
situation between themselves with input data sequence $X^A_n$ from A, an observed random  variable
$Y_n^B$ for B that leads to a better error performance than that
obtained by E from her observed random variable $Y_n^E$ and all her
side information.

\item \emph{Error Correction}:

The users agree on a generated string
that is free of error with high probability if E is absent.

\item \emph{Privacy Distillation}:

The users derive from the generated
string a generated key $G$ on which E's
error probability profile satisfies a given security level.
\end{enumerate}
\vspace{5mm}
The index $n$ above measures the number of channel output uses. In a quantum protocol, $Y_n^B$ and $Y_n^E$ are obtained from quantum measurements on the quantum signal space accessible to B and E. The term ``advantage distillation'' has been used previously \cite{gisin02} for the situation in which the above advantage is created by postdetection selection of data by B. That is one possible way to create advantage classically as described by Maurer \cite{maurer93} and the Yuen-Kim protocol \cite{yk98}. Note that a shared secret key between the users is needed for this approach with public discussion exactly as in BB84 type protocols, for message authentication during key generation to thwart man-in-the-middle attack.

Eve's \emph{conditional probability distribution} (CPD) is the probability distribution of the different possible data values that she would obtain by processing whatever is in her possession. In the case of classical continuous signals, say a real-valued random vector $y_n^E$  which E observes, she can obtain from this $y_n^E$ the different probabilities $p_i, i \in \{1, \cdots, N=2^n\} \equiv \overline{1-N}$ for the possible $N$ $n$-bit data sequences $x_i$ that A transmitted via the signal. Indeed, $p_i \equiv  p(x_i|y_n^E)$, which can be computed from the {conditional} probability $p(y_n^E|x_i)$ and the data a priori probability $p(x_i)$.

In the quantum case, a measurement has to be first selected by Eve on her probe or copy with result $Y^E_n$. The $p_i$'s are obtained accordingly where now $p(y_n^E|x_i) = \mathrm{tr} \Pi_{y_n} \rho_{x_i}$ where $\{\Pi_{y_n}\}$ constitutes the measurement PO(V)M \cite{ykl75,helstrom76} and $\rho_{x_i}$ is the $x_i$-dependent state in Eve's possession. Note that Eve's CPD $\{p_i\}$ is indeed conditional not only on all the relevant system parameters, but \emph{also} on her specific measurement result $y_n^E$.

In the above privacy distillation step, classical processing is used by B to distill ${G}$ from $Y_n^B$. On the other hand, Eve obtains from $Y_n^E$ an estimate $\hat{G}$ of $G$ with a whole CPD $\{p_i\}$ on all its possible values. The term ``privacy amplification'' is standard in the QKD literature when the Shannon or Renyi entropy criterion is used to measure how well $\hat{G}$ approximates $G$ \cite{bennett95}. In Section III.D it will be shown that a {necessary} criterion is $p_1$, Eve's optimal probability of getting $\hat{G}=G$. In either case and in general, ``privacy'' cannot be ``amplified'' but only distilled or concentrated by processing, exactly as in the case of quantum entanglement distillation. However, as in the case of the term ``QKD'', we will use ``privacy amplification'' to denote the usual privacy distillation procedure and sometimes even all such, for convenience. Note that steps (ii) and (iii) are often combined in a single extraction step, as in the case of many QKD security proofs, though conceptually and in concrete implementations they are distinct steps and goals.

We give the usual entropy description of advantage creation before discussing the more appropriate CPD one. The situations described by information measures instead of probabilities are often quantitatively meaningful only asymptotically, except in the all or none limits. Thus we omit the $n$-dependence and use $X_A$, $Y_B$, $Y_E$ to denote the data chosen by A and the observations B and E make. Advantage is created in entropy terms if and only if the conditional entropies obey $H(X_A|Y_E)<H(X_A|Y_B)$, or in terms of mutual information
\begin{equation} \label{advantagecreationcondition}
I(X_A;Y_E) < I(X_A;Y_B).
\end{equation}
This was first used to propose key generation with the ``wiretap channel'' of Wyner \cite{wyner75}, later generalized as a general condition \cite{ck78} and relaxed by Maurer when (authenticated) public discussion is included \cite{maurer93}. It may be observed that (\ref{advantagecreationcondition}) still holds in the latter case if $Y_B$ is taken to be the post-selected values. Thus, it is appropriate to consider (\ref{advantagecreationcondition}) as the advantage creation (for B vs E) condition when $X_A$, $Y_E$, and $Y_B$ are appropriately selected from a protocol. It has clear intuitive meaning and mathematical significance as the condition on secrecy capacities \cite{ck78,maurer93} of the appropriate ``channels''.

When a shared secret key $K$ is utilized between A and B for the key generation, the situation can in general be represented
\begin{equation} \label{encryption}
y_n^B = \mathcal{E}(x_n^A, K ,r),
\end{equation}
where $\mathcal{E}$ is the encryption map (including fixed channel on data transmission) that includes a randomizer $r$ that is not known to B and may not be even known to A such as when $r$ is some system noise. Unique decryption means
\begin{equation} \label{decryption}
x_n^A = \mathcal{D}(y_n^B,K)
\end{equation}
for an openly known function $\mathcal{D}$ which would yield the correct $x_n^A$ without knowledge of $r$. We have described the situation of ``randomized encryption'' \cite{yuen05,nair06} where (\ref{decryption}) can be expressed as
\begin{equation} \label{entropydecryption}
H(X_\mathbf{A}|Y_\mathbf{B},K)=0.
\end{equation}

Can a shared secret key $K$ be used between A and B to generate $G$ with $H(G|Y_\mathbf{E}) > H(K)$? This inequality is required for fresh key generation from an ``information theoretic'' security point of view so that $G$ is more than just merely $K$ in another guise. We assume no public discussion which requires a message authentication part with shared secret key to complete the protocol. The usual Shannon limit on data encryption \cite{yuen05,yuen06,nair06}, where $\mathcal{E}$ is known but the arguments of (3) are unknown  to Eve, is given by
\begin{equation} \label{shannonlimit}
H(X_A|Y_E) \leq H(K).
\end{equation}
 Condition (\ref{shannonlimit}) says that, given (3)-(5), there is no more (entropic) uncertainty on $X_A$ than the key $K$ itself, assuming $\mathcal{E}$ and $\mathcal{D}$ are openly known. Fresh key generation is possible {with cost $\mbf{K}$} only when
\begin{equation}
I(X_A;Y_E) < I(X_A;Y_BK) \mathbf{-} H(K),
\end{equation}
 which is the same as
\begin{equation}
H(X_A|Y_E) > H(K)
\end{equation}
under (5). In arriving at (7) it is assumed that the key $K$ is not to be used in any other way so it has to be subtracted in counting how many fresh key bits are generated. {Key generation is impossible when $Y_E=Y_B$ from (6) and (8)}.

\subsection{Security Measure and $p_1$}
Let us consider the issue of security measure on the generated key
$G$ in a  key generation system. Eve's Shannon
entropy $H_E(G)$, or equivalently her mutual information
$I_E=|G| - H_E(G)$, is the most commonly used measure. If
Eve's knowledge of $G$ is bit by bit, the binary entropy of a
bit is in one-one correspondence with Eve's bit error rate.
However, exactly as a many-body problem in physics, in general Eve has bit-correlated information on $G$,
and we may ask: What is the concrete security guarantee provided
by having $I_E \leq \epsilon$ for a given level $\epsilon$? The
problem arises because $I_E$ or $H_E$ is a theoretical quantity
with no operational meaning automatically attached. In standard
cryptography, this issue does not arise because fresh key
generation is considered impossible \cite{yuen03,yuen05,yuen06,nair06}and was never attempted,
while security of other cryptographic functions is based on
computational complexity.

In ordinary communications, the operational significance of the
entropic quantities is given through the Shannon source and
channel coding theorems, which relate them to the empirical
quantities of data rate and error rate. But what is the
corresponding empirical security guarantee in cryptography? This
issue was \emph{not} addressed by Shannon in his classic
cryptography paper written at about the same time as his classic
information and communication theory papers. It was not addressed
by anybody else since, except briefly in \cite{yuen06qcmc}.

The general situation of security guarantee on a data string is as follows. The attacker could derive a probability estimate $\{p_i\}$, her CPD, on the $N=2^n$ possible $n$-bit strings as described in the previous section. This $n$-bit string could be the generated key $G$ from a key generation protocol, or the data $X_n^A$ in a direct encryption system. If Eve knows nothing about the string, $p_1 = p_2 = \cdots = p_N = 2^{-n}$ which is \emph{uniform} randomness. Also, \emph{any} subset of $m$ bits from the $n$-string, $1 \leq m \leq n$, has a probability $p(m)=2^{-m}$ for her. Thus, Eve has no information at all on the string.

Any quantitative security measure one adopts must be a function of $\{p_i\}$. A one-number numerical function of the $\{p_i\}$ (that does not encode the whole CPD) may not capture the different $p_i$ values in the CPD in general, other than the extreme limits of uniform randomness and being nonrandom. It would express \emph{merely a constraint} on the possible $\{p_i\}$. In particular, Eve's probability of successfully getting any subset of the $n$ bits correct is determined by $\{p_i\}$. The \emph{important} point is that one must know that the measure adopted actually captures the security feature one desires in an empirical operational sense.  In the following and in Appendix A, we will demonstrate that the entropy measure $I_E$ cannot do a good job in general. In Appendix B we will show the variational distance $\delta_E$ between Eve's CPD and the uniform distribution $U$ {is much better but still numerically inadequate}.  We will suggest that $p_1$, Eve's maximum probability, is a {necessary} one-number measure for key generation protocols while {\emph{still}} being far from adequate. It appears that Eve's total probability profile to be described later {or at least a combination of $\delta_E$ or $I_E$ with $p_1$ is needed for proper security guarantee}.

First of all, the CPD and in particular $p_1$ has the same clear operational significance as probability. For a meaningful security guarantee, $p_1$ {\emph{must}} be sufficiently small. For a moderate length $|G| \sim 10^2 - 10^3$, one may argue that $p_1 \sim 2^{-20}$ is not small enough for some applications while $2^{-10}$ would be a disastrous breach of security because {it is possible} the whole $G$ could be found by Eve with a probability of $0.001$. The natural question on the adequacy of the entropy measure is: Assuming $H_E(G) \geq m$, what is the worst possible $p_1$ from the security viewpoint among the possible $\{p_i\}$ that satisfies this condition. It turns out that the case where E knows $m$ bits out of the $n$ exactly is in a sense the best possible security, \emph{not} the worst. Note that there is \emph{no} meaning to average over the possible CPD under a fixed $H_E(G)$ to get an average $p_1$. One CPD is already fixed by the system and the attack, and that is the only correct one to use. If only $H_E(G)$ is known, it is \emph{not} guaranteed that the biggest possible $p_1$ would obtain with only a small probability.

For fixed $p_1$, the largest $H_E(G)$ among all CPD has the remaining $1-p_1$ uniformly spread the other $N-1$ possibilities.  It is given by $H_E(G) = \max_{p_1} F(p_1)$ with
\begin{equation} \label{F}
F(p_1) = h_2(p_1)+ (1-p_1) \log_2 (2^n -1),
\end{equation}
where $h_2(\cdot)$ is the binary entropy function. There is no need to maximize over $p_1$. It is easy to show from (9) the following \\ \\
\textbf{Lemma 2:}\newline
 $p_1 \geq 2^{-l} - [\frac {n}{\log_2(2^n -1)} - 1]$  \hspace{3mm} for \hspace{3mm} $I_E/n \leq 2^{-l}.$\\ \\
Since $\frac {n}{\log_2(2^n -1)} - 1 < \frac{1}{n2^n},$ it is very small compared to $2^{-l}$ for any $l \leq n$ and moderate $n$. Thus, we have
\begin{equation}
p_1 \sim 2^{-l} \hspace{4mm} \textrm{for} \hspace{4mm} I_E/n \sim 2^{-l}.
\end{equation}

If a constraint on $p_1$ is first imposed instead, $p_1 \leq 2^{-l}$, the smallest $I_E$ is given by (10) while the information limit on E is only a weak guarantee as given by the following \\ \\
\textbf{Lemma 3}:
\begin{equation}
H_E(G) \geq l \hspace{4mm} \textrm{for} \hspace{4mm} p_1 \leq 2^{-l}.
\end{equation}
\textbf{Proof}: From the (Schur) concavity of $H$, the minimum $H_E$ under given $p_1$ occurs at $p_1 = \cdots = p_m =2^{l}$ for $m=2^l, p_{m+1}= \cdots = p_N =0$. \\

Lemma 3 tells the obvious fact that under $p_1 = 2^{-l}$, all but $l$ bits of $G$ could be completely known to Eve. This does \emph{not} suggest that $p_1$ is not a good measure for key generation, because in a specific problem such as the ones in section II and section VII, the other $p_i$ are either explicitly known or readily estimated. In particular, it is usually clear that there are many not totally known bits in $G$. However, it is also {clear that $p_1$ by itself is not generally sufficient and it is} useful to have a bound on $I_E$ that would rule out this disastrous possibility.

Generally, if Eve can try $m$ different possible $G$ to break
the cryptosystem, the first $m$ $p_i$ are the relevant numbers to
determine any quantitative level of security. For $N$ possible
trials, the trial complexity $C_t = \sum_{i=1}^{N} i\cdot p_i$
which is the average number of trials Eve needs to succeed, is a
meaningful measure of security. Note that in this definition of $C_t$, Eve already follows the optimal strategy of testing the more probable sequences first according to the order $p_1 \geq \cdots \geq p_N$. We have, similar to Lemma 3, \\ \\
\textbf{Lemma 4}:
\begin{equation}
C_t \geq (2^l +1) /2 \hspace{4mm} \textrm{for} \hspace{4mm} p_1 \leq 2^{-l}.
\end{equation}

In this connection, it may be pointed out that information theoretic  security could be \emph{no} better than a complexity measure if many trials are allowed. In particular, an $n$-bit uniform uncertainty {can be removed with} no worse than $2^n$ possible trials, or $2^{n-1}$ on average.

In Appendix A we will discuss how exponentially small $I_E/n$ of (10) is not a good security guarantee unless $l \sim n = |G|$. For a finite-$n$ concrete cryptosystem, it appears difficult to approach this experimentally while it may be easier to show that  for $p_1$.

It may be mentioned that $p_1$ is equivalent to the $H_\infty$ entropy often used in statistical analysis. However, its significance for the characterization of random bits sequences in key generation has not been spelled out as done in this paper.

\subsection{Uniform $\epsilon$-Random Bit String {and Variational Distance}}

A general security measure on $\{p_i\}$ would include not just $p_1$ but also the probabilities of various subsets of the bits in $G$, which are \emph{not} determined by $p_1$. Some measure on the closeness of these probabilities to the uniform $U$ is needed.

To appreciate the importance of such probabilities, consider the following probability distribution on the $n$-bit $G$. One subsequence, say the first $m$ bits, $1 \leq m \leq n$ , occurs with a probability $p$ independently  of the rest. Assuming the rest is uniformly distributed, we have for $L=2^{n-m}, \sum_i q_i=1$, the following distribution on the $2^n$ possible values of $G$:-
\begin{equation}
pq_1, \cdots, pq_L, \frac {1-p}{2^n - 2^{n-m}}, \cdots, \frac {1-p}{2^n - 2^{n-m}}.
\end{equation}
Under the constraint $I_E/n \leq p_1$ for given $p_1$, it follows from (13) that Eve could determine the first $m$ bits with a probability
\begin{equation}
p \sim \frac {n}{m} p_1
\end{equation}
assuming $2^n \gg 2^m$ and $p_1 \leq m/n$. Equation (14) shows that a smaller subsequence of $G$ may possibly be determined with higher probability than the maximum $p_1$ of the whole $n$-bit sequence, in linear proportion to its size. Its possible disastrous effect on security is illustrated numerically in Appendix A.

One useful measure that would yield meaningful bounds on the subsequence probabilities of $G$ is uniformly $\epsilon$-randomness. A \emph{uniformly $\epsilon$-random} $n$-bit string $G$ is one for which
\begin{equation}
|\mbf{p}_i - \frac{1}{N}| \leq \epsilon_n ,\hspace{10mm} N =2^n.
\end{equation}
That is, all the probabilities of the different sequences deviate from the uniform probability $1/N$ by at most $\epsilon_n$. A more manageable single-number criterion similar to (15) may be used, the usual variational distance $\delta(G,U)$ between $G$ and the uniformly distributed $U$,
\begin{equation} \label{variationaldistance}
\delta(G,U)= \frac {1}{2} \sum_{i=1}^{N}|p_i - \frac {1}{N}|.
\end{equation}
It can be readily shown that if $\delta(G,U) \leq \epsilon$, the probability $p$ of getting any $m$-bit subsequence correctly is bounded by $p \leq \epsilon + \frac {1}{2^m}$.

\subsection{Privacy Amplification and $p_1$}
There is another great significance of $p_1$ on $G$ -- it determines the length of a uniformly random string that can be extracted from $G$ by privacy distillation of any kind. {This is relevant if $G$ is the bit string B has before privacy distillation instead of the final generated key.} For $p_1 = 2^{-l}$, \emph{no distilled key $\tilde{G}$ can be obtained from $G$ which has $p_1 < 2^{-l}$}. This is because distillation is obtained by an openly known map $D: \{0,1\}^n \rightarrow \{0,1\}^m$ that maps the $n$-bit $G$ to an $m(<n)$ bit $\tilde{G}$. If another secret key is used in this process,  its randomness uncertainty has to be {counted} also. Privacy cannot be ``amplified'', it can only be concentrated in a shorter key within the $p_1$ limit. Note, however, that the subset probabilities $p$ of (13) may be improved by privacy distillation, especially for $m/n$ small.

In the case of statistically independent bits in $G$, say probability $p_0 > 1/2$ for Eve to correctly obtain a bit, $p_1 = p_0^n$ for an $n$-bit $G$. Thus $p_1 = 2^{-l}$ for $l=\lambda n$ for $p_0 \equiv 2^{-\lambda}, \lambda <1$. In this case, privacy amplification on $G$ can be used to produce even a nearly uniform $\tilde{G}$ with a linear rate $\lambda$. {In the more general statistically dependent case, there is no known result that would guarantee the input entropy per bit of a privacy amplification code is increased at the output. A different criterion is used in \cite{renner08} and will be discussed elsewhere. Note that the above} $p_1$ limitation shows that in general $I_E$ \emph{cannot} be made small exponentially {by privacy amplification} beyond a fixed limit given by (10). The prevalent contrary impression that there is no such limit is \emph{incorrect}.

Even within the limit $p_1 \leq 2^{-l}$, there appears no known algorithm that would compress an $n$-bit $G$ with an arbitrary CPD to an $l$-bit $\tilde{G}$ with a prescribed near uniform {distribution}. Indeed, the mere possibility of such distillation is unknown and appears to be a useful and promising area of research.

\subsection{Key Generation via $I_E$ and $p_1$}
We will describe schematically how a key generation scheme may be obtained under (2) and the $p_1$ criterion. With the usual ``capacity condition'' (2), the users can choose a data transmission rate $R$ that satisfies, similar to (1),
\begin{equation}
I(X_A;Y_E) < R < I(X_A;Y_B).
\end{equation}
If the above $X_A$, $Y_B$, $Y_E$ could be carried out many times in a statistically independent fashion as in the case of memoryless channels, a key generation scheme can be specifically obtained as follows. Alice picks a code with rate $R$ for transmitting the data that satisfies (17) that B can decode in practice. From the Shannon channel coding theorem \cite{gallager68}, {Bob's} error probability can be made exponentially small in $n$, the number of channel uses. From the ``Strong Converse'' to the Coding Theorem \cite{viterbi79}, Eve's error probability $1-p_1$ is bounded by
\begin{equation}
p_1 \leq e^{-nE_s(R)},
\end{equation}
for an exponent $E_s(R)$ that can in principle be evaluated for given $p(X_A|Y_E)$. {For the qb-KCQ scheme of section II, the exponent $E_s(R)$ in (18) for collective attacks can be explicitly evaluated and will be presented elsewhere.} This $E_s(R)$ gives a linear key generation rate which is nonzero when $R>I(X_A;Y_E)$. As discussed in the last subsection, hopefully an $n$-sequence $Y_n^B$ can be compressed to an $nE_s(R)$-sequence that is nearly uniformly random.

This generalizes to the finite $n$ case without $I$'s as follows. Let $X_n^A$ be sent in a coded/modulated system so that it can be recovered with sufficiently small error probability via $Y_n^B$. Let the optimal error $1-p_1$ that Eve can obtain on $X_n^A$ satisfies $p_1 \leq 2^{-l}$. Then an $l$-bit $G$ could be generated between A and B from an algorithm that compresses $X_n^A$ to a nearly uniform $l$-bit string. Since there is no chance for E to encode the data $X^A_n$, a detection theory formulation for her CPD $\{p_i\}$ is more appropriate than an entropic one. It may yield a more favorable bound on $p_1$ from the users' viewpoint than the generally applicable (18).

As discussed in appendix A, Eve's $p_1$ actually depends on the observed $y_n^E$. In this section III, we have talked about $p_1$ as if it is unique independent of $y_n^E$. While such a situation may obtain in a protocol such as that of section VI, that is a rare exception and not a rule. Thus, the $p_1$ of (18) is actually the average $\overline{p}_1$ of $p_1(y_n^E)$ over all the $y_n^E$, as is the $p_1$ obtained in the usual classical and quantum detection as well as communication theory. {The use of Markov's inequality for a nonnegative-valued random variable may allow such $\overline{p}_1$ to be used with a more stringent requirement of $\overline{p}_1 < \epsilon^2$ compared to the original prescription $p_1 < \epsilon$. This follows from \cite{cover91}
 \begin{equation} \label{markovinequality}
 \mathrm{Pr}[X \geq \delta] \leq E[X]/{\delta}
 \end{equation}
 with $X$ being $p_1(Y_n^E)$ that is conditioned on Eve's observation.} A sufficiently small such averaged $\overline{p}_1$ is clearly \emph{a necessary condition for security}.

\section{KCQ and Key Generation via Secret Key}
This section first describes the basic issues in key generation during which a shared secret key is employed. Unfortunately, a shared secret key is needed in all known key generation protocols, classical or quantum. {The} reason for this {in BB84 type protocols} is that the users need to thwart man-in-the-middle attack {since} there is ``public'' exchange in the protocol. In this attack, Eve intercepts the communication line and pretends to be A while exchanging with B to set up the key agreement, and pretends to be B while exchanging with A. She intercepts the subsequent communications when A and B use their generated key, and obtains full information without being detected. It is not sufficient that A and B authenticate themselves outside the times of protocol execution. Eve could attack only during such times. As in other cryptographic protocols \cite{menezes97}, man-in-the-middle attacks have to be dealt with by the protocol itself. One way to do that is to employ a shared secret key for message authentication to detect such attacks during protocol execution. It has not been treated quantitatively as part of any QKD protocol thus far. We will show that it is a serious issue the significance of which is yet to be assessed.

A shared secret key is {in some sense} used in an even more essential way in KCQ key generation. We will describe the schematics of such quantum key generation in a full generic protocol and the significance of various  security {assumptions} that can be meaningfully employed.

\subsection{Problem of Key Generation with a Shared Secret Key}
The major issue in key generation with a shared secret key $K$ is that Eve can launch an attack with a guessed value of $K$, which will be called the \emph{guessing attack}. In protocols with public discussion including all BB84 type protocols, Eve could guess at $K$ and succeed in breaking the system completely with $p_1=2^{-|K|}$. In a KCQ protocol such as the qubit protocol of Section II, Eve could make the measurement on the signal just out of A's transmitter that corresponds to a chosen value $k$. Again she would succeed completely for any $n$-segment with a probability $p_1=2^{-|K|}$.

{This guessing attack may be considered a generalization of the Shannon limit (6) applicable to both classical and quantum cryptography. In the quantum case, one may try to get around it by weakening the meaning of ``fresh key generation''. The BB84 and KCQ approaches may be considered as two very different ways to deal with this limit.}

In view of our treatment in Section III, what could be the meaning of key generation in {this} situation with the claim that its length is greater than the $|K|$ necessary for a fresh key? The situation remains the same is if one averages $p_1$ over all possible $K$ values. Indeed, $p_1=2^{-|K|}\cdot 1 + \cdots \geq 2^{-|K|}$ with the averaging.

Various qualifications on the security claim can be made to allow for ``fresh key generation'' in some sense. However, it is clear that the $n$-bit key $G$ (or $\tilde{G}$) generated is \emph{not} the same as an ordinary shared secret key from which $G$ cannot be obtained with probability $2^{-l}$ for $l<|G|$. We would discuss the situation of KCQ in the following. As to BB84 type protocols, it may be observed that depending on how the exact message authentication method is used in the protocol, Eve may be able to combine the guessing of a subset of the $|K|$ bits or some other attack on the message authentication with her quantum attack and obtain information beyond what is quantified in the literature. This is the case regardless of what security measure is used including $I_E$, but {does not appear to have} been dealt with in the literature.

Note that $p_1$ and $I_E$ together, say in the form of (10) , rules out the possibility that the generated key $G$ can be obtained from classical key expansion \cite{note2}. This can also be guaranteed from $p_1$ obtained via a quantum or classical meausurement by an attacker on the cryptosystem as discussed in {Section III}. It seems a clear meaning on key generation can only be obtained if one has the \emph{total probability profile}  which gives Eve's CPD for the $2^{|K|}$ different $K$ values under any specified attack. {On the other hand}, since the guessing attack is only good at probability $2^{-|K|}$, one  may consider that ``satisfactory'' for moderate $|K|$ regardless of the length $|G|$ generated. This would make public exchanged protocol ``secure'' and also rule out the correct $K$ guessing attack in KCQ protocols as relevant.

\subsection{General KCQ Key Generation}
Consider an entire joint process of
data transmission and encryption/decryption as described in Fig. 2.
A sends an $m$-bit sequence $U_m$ and encrypt/encode it into
an $n$-qubit or $n$-qumode
 sequence in state $\rho^k_x$ with the possible use of
a shared secret key $k$ with B,
which may include a source code key, a channel code key ,
and a quantum state modulation code key.
Classically, $\rho^k_x$ would be replaced by just an $n$-bit channel
input sequence $X_n$ corresponding to the $x$ in $\rho^k_x$.
The `channel'
represents all the interference from the system one has to suffer,
with $Ch^i$ giving output  states for $i=$ E, B. For E who does
not know $k$, the state is $\tilde{\rho}_x^E$ upon which she picks
a measurement on the basis of that and
her later knowledge from all sources
including public discussion to produce an estimate
 $\hat{G}$ of $G$, the final key generated by A and B.
 For B who
knows $k$, the channel output state is $\tilde{\rho}^k_x$ from which
she uses her knowledge of $k$ to obtain an estimate of
$\hat{U}^B_m$ of $U_m$. Classically, the states would be replaced by
 the
observations $Y^E_n$ and $Y^B_n$, the disturbed output of $X_n$.
Quantum mechanically, they are the results of measurements made on the qubits or qumodes from
which the estimates $\hat{U}^B_l$ are made. One may
consider that $Y_n^E$ is obtained without
knowledge of the modulation key.
Privacy distillation may already be incorporated in this process,
or may be added to $U_l$ and $\hat{U}^B_l$.
\begin{figure*}\begin{center}
\rotatebox{-90}{\includegraphics[scale=0.7]{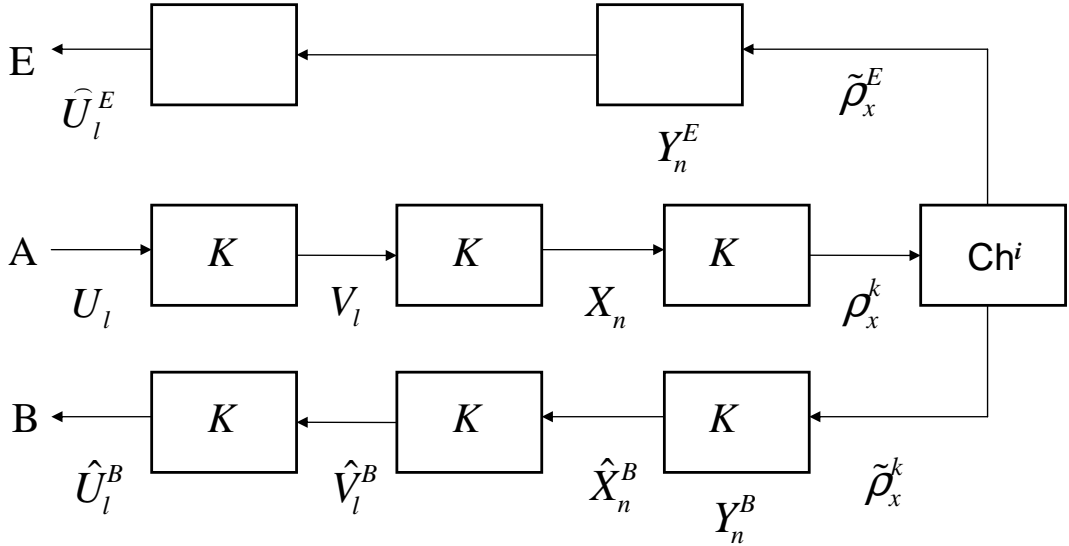}}
\caption{General keyed communication in quantum noise.}\end{center}
\end{figure*}
The essential steps in the operation of a KCQ key generation protocol
involve

\vspace{5mm}
(1) The use of a shared secret key $K$ between
A and B that determines the quantum states generated for the data
bit sequences in a detection/coding scheme between A and B that
gives them a better error performance over E who does not know
$K$ when she makes her quantum measurement;

(2) A way for
A and B to extract a fresh key from the above performance
advantage;

(3) A key verification process between A and B.
\vspace{5mm}

 The main novelty and power of
this approach, in principle, consists of

\vspace{5mm}
(a) Performance advantage is derived from the different quantum
receiver performance between B who knows the key $K$ when
she performs her quantum measurement and E who does not know $K$ when she makes her quantum measurement.

(b) No intrusion level estimation or even intrusion detection is needed
by A or B.

(c) No public discussion is needed between A and B.

(d) No separate privacy distillation, or reduction in the key
generation rate due to any such equivalent operation, is needed in
a properly designed system. \vspace{5mm}

As a consequence, this approach makes possible the development of
an efficient key generation protocol over long-distance
telecomm fibers using commercial optical technology.

A final \emph{key verification} step is needed in KCQ protocols. For the purpose of assuring the same key is agreed upon for future use, this step is recommended for all key generation protocols including BB84. For KCQ protocols, where no {intrusion level estimation} is carried out, it is needed to make sure that E has not messed up the key generation process so that A and B have different versions of the generated key $G$. This verification step can be achieved by any message authentication method \cite{menezes97} including ones that are not keyed. There can be no man-in-the-middle attack in KCQ protocols because {there is no public exchange and} Eve cannot get $G$ from A without knowing $K$. Eve could not tell more about $X_A$ other than what she could find out from her copy. Her disruption may lead to different versions of $X_A$ for A and B, which is to be detected by the verification step. If she disrupted, but A and B still get the same agreed key, she does not know anything more about {the generated key anyway,} in contrast to protocols that involve public exchange.
Schematically, a complete KCQ protocol corresponding to the communication
situation of Fig.~2 may be summarized as follows.

\vspace{5mm}
{\it Generic KCQ Protocol:}

(i) A picks a random bit sequence $U_m$, encodes and modulates the
corresponding $n$ qubits or $n$ qumodes as in Fig. 2, with a total
secret key $K$ shared with B.

(ii) From $K$, advantage creation is achieved via the different error
performance obtainable by B and E who does and does not know
$K$ at the time of their quantum measurements.

(iii) Privacy distillation may be applied to generate a net key $G$ on which E has an error probability profile that satisfies the security goal.

(iv) A and B verify that they agree on a common $G$.
\newline\newline
{Note that $|K|$ bits have to be subtracted from the generated key in the key generation rate. A net fresh key still results in the situations of sections II, V-VI where a linear key generation is obtained for a fixed $K$ under constant measurement attacks. In general, it is part of the performance/security analysis to ascertain the efficiency of key generation.}

\subsection{Security Approaches for KCQ Key Generation}

In analyzing the security of KCQ key generation schemes, we typically grant a full copy of the quantum signal at the transmitter to Eve for the purpose of bounding the information and performance she could possibly obtain in any attack. We did it in Section II on qubit key generation. Realistically, Eve may or may not be able to obtain such a full copy. In the process of doing so, say in the qubit case, she may introduce large errors that would prevent a key from being generated and such failure would be detected in the key verification step. On the other hand, for coherent-state signals in the presence of {large} transmission loss, she could actually accomplish that physically {with no} disruption to the protocol. It does not really matter which is the case from our security analysis viewpoint, as we are merely bounding her achievable performance with this ploy.

A question arises to whether Eve is supposed to know the shared secret key $K$ at some later time. If she does and she has sufficient quantum memory, the generated key $G$ would be completely compromised under the above ploy of granting her one full copy of the quantum signal. If there is not sufficient quantum memory, which is surely the case for at least the intermediate future as no realistic quantum memory of just $1$ sec long is even in sight, she would have to make a quantum measurement before she knows the key. Under such a situation, key generation is possible unless she launches a key guessing attack and hits on the correct $k$ value.

It is our contention that there is {little} reason to worry about the case where Eve would know $K$ at any time. Having a presumably secret key betrayed is an altogether different problem that occurs in every situation involving such secret. In the following, we will {just} use this as an additional ploy to bound Eve's performance. As will be seen, its use leads to a realistically useless bound for binary detection {in section V} but still a very strong bound for the $N$-ary CPPM of Section VI.

It may be observed that the situation is different with respect to the message authentication key in BB84 type protocols. If that key is not known during protocol execution, its knowledge is useless after the key is generated even with indefinite quantum memory. But as we remarked, there is no reason why Eve would know $K$ ever. It may be emphasized in this connection that there is no known-plaintext attack on the key in key generation. The data are secretly chosen by A with no regard to inputs from others.

Other than the guessing attacks, Eve's optimal quantum joint attack on the data could be formulated as follows. Let $\rho_x^k$ be the quantum state for a $K$-value $k$ and $X^A$ value $x$, thus $\rho_x =\sum_k {p_k \rho_x^k}$. The optimal quantum detector that leads to $\overline{p}_1$ for Eve is an $N$-ary digital detection problem, $N=2^n$. If the resulting $p_1$ can be upper bounded in the form $p_1 \leq 2^{-l}$, the possibility and meaning of key generation has been discussed in Section III. Similarly, one may consider individual or collective constant quantum measurement attacks, in which a reasonable measurement is chosen for each qubit or qumode in the signal state space, and find the optimal joint classical detector performance from such measurement results. They may be regarded as the correspondents of joint and collective attacks on BB84. In view of the \emph{great empirical difficulty} of measurement across more than one or two modes, the  resulting key generation thereby has \emph{clear practical significance}. In my view, its significance is even greater than that of a more general security analysis that is based on highly idealized model that never corresponds to reality. Further discussion on such issues are given in Section VIII.

In the following two sections, we would analyze the performance of coherent-state KCQ systems assuming a fixed measurement is made on each mode under collective attacks. Their performance under joint attacks are difficult to obtain and yet to be derived, but they are of great interest because these systems can be empirically implemented in a regime with much higher effective key generation rate than other QKD systems.

{Eve may attack the data via attacking the key first. A separate key security analysis has to be performed on each specific KCQ protocol. The key is perfectly secure for the qb-KCQ scheme of section II, and also for the CPPM scheme of section VI when $p_1$ is properly adjusted. Additional analysis on the binary scheme of section V is needed to tell the extent of modification required for key security.}

\section{KCQ Coherent-State key generation with binary detection}

In this section we describe the use of KCQ on qumodes,
quantum modes with infinite-dimensional Hilbert state spaces, for
key generation via coherent states of intermediate or large energy. The use of homodyne/heterodyne detection in quantum cryptography was suggested in \cite{yuen96nasa}, and in
conjunction with coherent states in \cite{yuen00}. In most of the
current experimental developments \cite{gisin02} of QKD, coherent
states are employed in BB84 type protocols that are limited in energy
to $\sim 0.1$ photon, if only because of the photon-number splitting
attack that E can launch near the transmitter \cite{yuen96,slutsky98}.  With
KCQ, we will in this and the next section show that much larger energy
can be employed, line amplifiers and pre-amplifiers can be used, and
conventional optical technology on the sources, modulators, and
detectors can be utilized. Furthermore, direct
encryption
coherent-state KCQ in what is called the $\alpha \eta$ scheme has
already been experimentally demonstrated
\cite{corndorf05,liang05}, which will integrate smoothly with the corresponding
key generation schemes.

The usual description of a single coherent state already involves
an infinite dimensional space, referred to as a {\it qumode}.
Similar to the qubit case in Fig.~1, we may consider $M$ possible
coherent states $|\alpha_l \rangle$ in a single-mode realization,
\begin{equation}
\label{l}
\alpha_l = \alpha_0 (\cos \theta_l + i \sin \theta_l),
             \hspace{3mm}
             \theta_l= \frac{2\pi l }{M},
               \hspace{3mm}
                l \in \{1,...,M],
\end{equation}
where $\alpha_0^2$ is the energy (photon number) in the state, and
$\frac{2\pi l}{M}$ is the angle between two neighboring states. In
a two-mode realization, the states are products of two coherent
states
\begin{equation}
\label{m}
 |\alpha_0 \cos \theta_l \rangle_1 |\alpha_0 \sin \theta_l\rangle_2
 \hspace{2mm}, \hspace{2mm}
             \theta_i= \frac{2\pi l}{M} , \hspace{3mm}
               l \in \{1,...,M\},
\end{equation}
The qumodes may be those associated with polarization, time,
frequency, or any type of classical mode. Any two opposite states on the circle form the basis states of a phase reversal keying (antipodal) signal set, which are nearly
orthogonal for $\alpha_0 \geq 3$. There are $M/2$ possible bases. The optimal quantum phase
measurement \cite{helstrom76,hall91} yields a root-mean-square phase error
$\Delta \theta \sim 1/\alpha_0$. Thus, on a bit-by-bit situation, when $M\gg \alpha_0$,
the probability of error $P^E_b \sim 1/2$ when the basis is not known
which has been confirmed numerically \cite{barbosa03}, while
$P^B_b \sim \exp(-\alpha_0^2) \rightarrow 0$ when the basis is
known.

The use of this scheme for direct encryption has been extensively studied theoretically \cite{yuen03,yuen06,nair06,yuen07} and experimentally \cite{corndorf05,liang05,barbosa03}. It is called $\alpha\eta$ or Y-00 quantum noise randomized scheme. It can be used for key generation as follows.

When the key is unknown to Eve, the general quantum measurement she could make in principle to cover \emph{all} possible signal sets is heterodyning {or phase measurement} on each qumode. Assuming this (or any other) individual attack, Eve could determine her whole CPD of $p(\mathbf{x}_n^A | \mathbf{y}_n^E)$ for the data $n$-sequence $\mathbf{X}_n^A$ sent by A with her {measurement} result $\mathbf{y}_n^E$ where each $y_i^E, i \in \overline{1-n}$ is a complex number. As discussed in Section IV.B, a whole copy of the signal is to be granted to Eve for obtaining this CPD for the purpose of security analysis. The presumably nearly optimal individual measurement for the $\alpha\eta$ signal set (20) or (21) for large $M$ is the optimal phase measurement. The best estimate of $\mathbf{x}_n^A$ from  $\mathbf{y}_n^B$ is a classical $N$-ary detection problem with $N=2^n$ that would provide Eve's best $\overline{p}_1$ of her CPD. Fresh key generation is possible if for some $n$, $p_1< 2^{-|K|}$ assuming it is essentially error-free for B, which may be achieved without coding as indicated above. However, no rigorous result has yet been obtained in this problem.

On the other hand, that fresh key generation must be possible under such attacks can be seen from the performance bound obtained by granting Eve the value of $K$ after the individual qumode measurements. In that situation, Eve could use the key value to solve the binary decision problem on each of the qumodes from each $y_i^E$ she got. In contrast, B could use the optimal binary quantum receiver or a close approximation thereof to determine the data bit of each qumode. For the discrimination of two equally likely coherent states
$\{|\alpha_0\rangle, |-\alpha_0\rangle \}$, the optimum quantum receiver yields
an error rate $\bar{P}_b$ that may be
compared to the heterodyne result $P^{het}_b$
and the phase measurement result $P^{ph}_b$, with $S= \alpha^2_0$,
\begin{equation}
\label{C}
\bar{P}_b= \frac{1}{4} e^{-4S}, \mbox{     }
P^{het}_b \sim \frac{1}{2} e^{-S}, \mbox{     }
P^{ph}_b \sim  \frac{1}{2} e^{-2S}
\end{equation}
Here, $S$ measures the average number of photons received in the
detector and (22) applies in the so-called quantum-limited
detection regime--- unity detector quantum efficiency, infinite
detector bandwith, all device noise suppressed. Under (22)
and dropping the factors in front of the exponentials for a
numerical estimate of the bit-error rate (BER), which is required
to be $\leq 10^{-9} $ per use in a typical communication
application, we have, for $S \sim 10, \bar{P}_b
\sim 10^{-12}, P^{het}_b \sim 10^{-3}, P^{ph}_b \sim 10^{-6}$. If
the data arrives at a rate of $1$ Gbps, the user B is likely to
have $10^9$ error-free bits in $1$ sec, while E would have $\sim
10^3 $ errors among her $10^9$ bits with the optimum phase
measurement. Presumably, the users can then generate $\sim 10^{3}$ secure key bits
by eliminating E's information. Thus, in principle, $\alpha \eta$
in its original form is capable of secure key generation against
collective attacks that employs the optimal phase measurement on
each qumode even if Eve knows $K$ afterwards.

There is an $N$-ary quantum detection problem for finding Eve's (averaged) $\overline{p}_1$ under joint attack,  {the performance of which would provide the security level under joint attacks.}

The $\alpha\eta$ key generation scheme in the form (20) or (21) allows a direct attack on its key by Eve similar to the case of direct encryption. This problem can be solved by additional randomization called DSR \cite{yuen03,yuen07}
which we would not go into here.

It may be mentioned that for binary detection of coherent states, the optimal quantum receiver performance cannot be better than that of heterodyning by $6$ dB in energy or error exponent. The antipodal signals of $\alpha\eta$ lead to {exponentially} optimal BER under energy constraint on binary coherent-state signals, which cannot be improved by bandwidth utilization \cite{yuen03}. The proofs of these statements will be omitted for brevity. This leads us to consider $N$-ary systems in the following. On the other hand, it should be emphasized that since (22) provides just a bound, presumably rather weak when Eve does not know $K$, there is much value in determining Eve's $p_1$ in such cases when the signals are moderately strong in the range $S \sim 10^3 - 10^4$ as in the experimental implementation of $\alpha\eta$ direct encryption.

\section{KCQ coherent-state key generation with $N$-ary detection}
The above limitation on the binary detection advantage of an optimal
quantum receiver versus heterodyne can be overcome in $N$-ary
detection. The use of $N$-ary systems, in fact, is one form of
coding. As will be seen in the following, it indeed corresponds to
driving the system at a rate between B's and E's mutual information
with respect to A as in (17). Amazingly, for the particular
CPPM (Coherent Pulse Position Modulation) system we now turn, such a rate choice by A can make $I_E$ go to zero with a flat error profile {and also with} (full)
information-theoretic security against known plaintext attack on the
key. This can be proved against the universal heterodyne attack, and is possibly
true against more general attacks.

An $N$-ary coherent-state pulse position modulation system has the
following signal set for $N$ possible messages,
\begin{equation}
\label{H}
|\phi_i\rangle = |0\rangle_1 \cdot\cdot\cdot
                   |\alpha_0\rangle_i \cdot\cdot\cdot
                   |0\rangle_N,
                   \mbox{          }
                   i \in \{1,...,N \}.
\end{equation}
In (\ref{H}), each $|\phi_i\rangle$ is in $N$ qumodes all of which
are in the vacuum state except the $i$th mode, which is in a
coherent state $|\alpha_0\rangle _i$. The corresponding classical
signals are orthogonal pulse position modulated if each mode is
from a different time segment, but generally the modes can be of
any type. For brevity, we retain the term `pulse position' even
through `general mode position' is more appropriate.

The photon counting as well as heterodyne error performance of
(\ref{H}) are well known \cite{gagliardi95}. The block error rate from direct
detection is exponential optimum for large $N$.
\begin{equation}
\label{I}
P^{dir}_e=(1-\frac{1}{N}) e^{-S}, \mbox{      }
\bar{P}_e \rightarrow e^{-S}.
\end{equation}
The optimum block error rate $\bar{P}_e$ for (23) is known
exactly \cite{ykl75}, and given by (\ref{I}) asymptotically.
 In contrast, for large $N$ the heterodyne block error rate
$P^{het}_e$ approaches $1$ exponentially in $n= \log_2 N$, which is
a general consequence of the Strong Converse to the Channel
Coding Theorem as discussed in section III.E.
For the present Gaussian channel case for heterodyne receivers,
explicit lower bound on the block error rate $P^{het}_e$, conditioned
on any transmitted $i$, can be obtained in the form
(p.~382 of \cite{gallager68}) that, for any $y$,
\begin{equation}
\label{J}
P^{het}_e > (1-[\Phi(y)]^n) \Phi(y-\sqrt{2S}),
\end{equation}
where $\Phi$ is the normalized Gaussian distribution. By choosing
$y > \sqrt{2n}$, (\ref{J}) yields explicitly
$P^{het}_e \rightarrow 1$ exponentially in $n$ for any given $S$.
It is a main characteristic of classical orthogonal or simplex signals
in additive white Gaussian noise that whenever an error is made, it is equally likely to be decoded by the optimal receiver to any of the $N-1$ other messages \cite{viterbi66}.
Thus, under the condition $P^{het}_e \rightarrow 1$, the CPD has  $p_i = 1/N$ for $ i \geq 2$.

The KCQ qumode key generation scheme CPPM works as follows.
Consider $N= 2^n$ possible $n$-bit sequences, and possible
coherent-states
\begin{equation}
\label{K}
|\psi_i\rangle = \otimes^N_{j=1} |\alpha_{ij}\rangle^{\prime}_j,
\hspace{5mm}  i,j \in \overline{1-N}
\end{equation}
in correspondence with $\{|\phi_i\rangle \}$ of (\ref{H}).
For simplicity, one may set $\sum_j |\alpha_{ij}|^2= |\alpha_0|^2= S$
for every $i$. Let $f_k$ be a one-to-one map between (\ref{H}) and
(\ref{K}) indexed by a key $K$. As an example of physical realization,
the connection between (\ref{H}) and (\ref{K})
could be through a set of $N$ beam-splitters with
transmission coefficients $\sqrt{\eta_m}$ for complex numbers
$\eta_m$, $m \in \overline{1-N}$, determined by $k$. Such a physical
realization combines the $\alpha_{ij}$ of (\ref{K}) coherently through
the $\eta_m$'s, and is represented by a unitary transformation
between the two $N$-tensor product state spaces
$\otimes^N_{i=1} H_i$ and $\otimes^N_{i=1} H^{\prime}_i$ for
the input and the output.
The states $|\psi_i \rangle$ of (\ref{K}) are used to modulate the
data $i$ by A, and B demodulates by first applying $f_k$ to transform
it to $|\phi_i \rangle $ of (\ref{H}) and then use direct detection
on each of the $N$ modes $H_i$.

Without knowing $f_k$ or $\eta_m$ so that there are both amplitude
and phase uncertainties for each $m$, it is expected that an attacker
can do very little better than heterodyne on all the $H^{\prime}_i$
 modes, which is equivalent to heterodyne on all the $H_i$ modes, and then
apply the different $f_k$'s on the classical measurement result. As presented above, by making $N$ large one can then
make not only $p_1 = 2^{-l}$ for any $l$ but E's error
profile is in fact nearly uniform, with $p_1 = (1-2^{-l})/(N-1) $
for $i \geq 2$ . This happens whenever $y_n^E$ leads to an error from the decision rule that minimizes the average error \cite{viterbi66}, which is asymptotically certain in the situation under consideration. Thus, if we choose the system parameters so that $p_1=p_i$, Eve would have uniformly random CPD's for all $y_n^E$ and $k$. As
a consequence, the system is not only completely secure against
ciphertext-only attack on the key but also fully secure against
known-plaintext attacks. There is no need for further privacy distillation. Also, in contrast to the binary detection case, the data is secure even if Eve has the key $K$ after her heterodyne measurement. {We summarize:}

{Against E's universal heterodyne attack, the $N$-ary CPPM KCQ
protocol can be made secure with key generation
rate $n= \log_2 N$ per use and uniform CPD to Eve.}

However, it is difficult to estimate $p_1$ closely and in the absence of such estimate, this one case difference among $2^n$ is either taken to be unimportant or additional DSR is needed to assure a fully uniform error profile.

The CPPM scheme is also ideal for direct data encryption because it
automatically produces (a near) uniform error profile on E.
Unfortunately, as in a classical orthogonal signaling
 scheme, large $N$ in CPPM means
exponential growth of bandwidth, not to mention growth in physical
complexity. Indeed, (24) itself is an infinite-bandwidth limit
result for large $N$. On the other hand, it is known \cite{yuen04}
that if the signal-to-quantum noise per unit bandwidth is small,
coherent-state direct detection systems do have larger capacity
than heterodyne ones. Thus, it may be expected that properly
designed error correcting codes, usually employed for bandlimited
systems for such reasons, could be developed to retain much of the
CPPM advantage for a large given bandwidth. {I would like to emphasize again that sections V-VI are sketchy introductions to some main ideas and possibilities of KCQ key generation with significant energy coherent states. Many details are yet to be developed.}

\section{Comparison with BB84}
We will briefly compare qualitatively KCQ key generation with BB84 type protocols which involve intrusion level estimation from a variety of viewpoints. No quantitative comparison will be attempted due to insufficient quantitative details in both cases on these issues.

\subsection{Unconditional Security}
It is often taken to be true that BB84 type protocols offer unconditional security in an information-theoretic sense, with at least asymptotic proofs supplied for the case of ideal devices. Some problems were raised concerning such proofs in \cite{yuen03} (App. A) which would not be entered into here. In any event, as discussed in ref. \cite{yuen06}, asymptotic existence proof has no practical implication in cryptology since one needs to analyze the security of specific finite-$n$ cryptosystems. Here, I would like to emphasize that quantitative information theoretic security of a bit sequence has yet to be made precise while being operationally significant. As discussed in Sections III, IV, and the appendices of this paper, the usual mutual information $I_E$ or variational distance $\delta_E$ or their quantum counterparts employed in the QKD literature is {not} a sufficient security guarantee except in an extreme region that appears to have little hope in ever getting realized practically. {Smallness of the} other measure $p_1$, the attacker Eve's optimal probability of getting the entire key generated, {is a necessary security feature.} Clearly a $0.1\%$ leak of $I_E$  is totally unacceptable -- See Appendix A. {On the other hand, no KCQ scheme security under joint attack has yet to be studied.}

Actually, as we further substantiate in the following, the practical significance of such ``unconditional security'' is over-rated, both because of intrinsic protocol modeling limitation and the futuristic technology granted to the attacker. It is not important to grant Eve the ability to have indefinite quantum memory while none of one second long is in sight. Similarly, there is no known experimental way to entangle three or more qubits or qumodes close to a given prescription. In comparison, it appears much easier for Eve to just break into a protected office to get some of the secrets by brute force.

Security under composition is another issue for both KCQ and BB84 assuming Eve has the required quantum memory. The purported solution for BB84 in ref. \cite{konig07} is {not valid as indicated} in Appendix B.

\subsection{Device Imperfections}

It has been well known that device imperfections could be exploited by Eve to seriously compromise a BB84 type protocol. Some seemingly irrelevant imperfections have been shown to be disastrous, beginning with the spectral defect of detectors pointed out in \cite{myers05} to the recent time-shift attack \cite{myers02,makarov06,qi07}
 that have been experimentally demonstrated \cite{zhao08}. In combination with the inevitable loss of an optical system, it was recently claimed that no loss more than a factor of $2$ can be tolerated for ``loophole free'' security \cite{ma08}. On the other hand, for KCQ key generation such as $\alpha\eta$ or CPPM discussed in sections VI-VII of this paper, there is no such sensitivity to device imperfection. The intuitive reason is clear: BB84 type systems operate with a very small signal level and thus are sensitive to small system parameter variations but KCQ may operate with much stronger signals.

\subsection{Sensitivity and Protocol Efficiency}
The performance of a key generation scheme for useful
real-life application is gauged not only by its security level, but also
its efficiency in at least two senses to be elaborated in the following.
For a protocol to be useful the \emph{two} efficiencies cannot be
too low.

The first type of efficiency that should be considered is {\it protocol
efficiency}, denoted by $P_{\mathrm{eff}}$, which has not been treated in the QKD
literature. It can be defined as the probability that the protocol is not
aborted for a given channel and a fixed security level
in the absence of an attacker E. It is essential to consider the
robustness of $P_\mathrm{eff}$ with respect to channel parameter fluctuation,
e.g., how {\it sensitive} $P_\mathrm{eff}$ is to small changes in channel
parameter $\lambda_c$ which may denote, e.g., the independent
qubit noise rate of any kind. In practice, $\lambda_c$ is known only
approximately for a variety of reasons, and imperfection in the system
can never be entirely eliminated. If $P_\mathrm{eff}$ is sensitive to such small
changes, the protocol may be practically useless as it may be
aborted almost all the time. Sensitivity issues are crucial in engineering
design, and there are examples of `supersensitive' ideal system
whose performance drops dramatically in the presence of small
imperfection. Classical examples include detection in nonwhite
Gaussian noise \cite{vantrees68}
and image resolution beyond the diffraction
limit \cite{goodman68}. Superposition of `macroscopic' quantum states is
supersensitive to loss \cite{caldeira85}. This crucial sensitivity issue is one of
fundamental principle, not mere state of technology. It has thus far
received little attention in the field of quantum information.

Our qumode KCQ key generation
protocols are robust to channel parameter fluctuations as the case of a conventional optical communication line. On the
other hand, e.g., the reverse reconciliation protocol in \cite{grosshans03}, which
supposedly can operate in any loss, is supersensitive in high
loss. Let $\eta$ be the transmittance so that $\eta \ll 1$
corresponds to the high loss situation. In the presence of a small
additive noise of $\eta/2$ photons in the system, the protocol
becomes insecure because the noise induced by the
attacker cannot be distinguished from excess noise. Note that high security
level often decreases $P_\mathrm{eff}$ and it is important to quantify
the tradeoff.

{Secondly}, even when the scheme is not supersensitive, the sensitivity level
has to {be} quantified in a QKD scheme involving intrusion level
{estimation in} a complete protocol with quantifiable security, for
the following reason that has {\it not} been discussed in the
literature. A stopping rule  for the protocol has to be adopted to stop the key generation process after it was aborted for a certain threshold number of times in a given time interval. If the  threshold is set too low, the protocol may be aborted too often by statistical fluctuation or un-modeled random disturbance and become inefficient. If it is indefinitely large, Eve may launch a very strong attack although it causes much disturbance. In any case, Eve could raise her possible information by counting on the users' repeated trials and launch a stronger attack than otherwise. A complete quantification cannot be obtained without an explicit stopping rule. {Such a rule would affect the quantitative efficiency of the protocol.}

\subsection{Effect of Loss}

The usual linear loss is extremely detrimental to quantum effects \cite{yuen04} and is also difficult to handle in physical cryptosystems. Eve should be presumed to be able to attack much closer to the transmitter than {Bob} at the users' receiver. In protocols with intrusion level estimation, it is customarily assumed that Eve could replace the resulting transmission link with a lossless one, the reason being that she could utilize free-space lossless links instead of fibers. In KCQ protocols without {intrusion level estimation}, Eve may gain a large energy advantage compared to {Bob} which has to be exceeded for fresh key generation.

Even for ideal devices it is not clear what kind of security proof has been supplied for a purely lossy BB84 system for what kind of specific protocol. Is the so-called ``twirling'' needed for security? Against what kind of channel replacement attack? We have these questions not just for coherent-state systems but also single-photon ones, which are not answered by the decoy state technique \cite{hwang03prl}. We also have these questions just according to the usual security measures adopted, not including $p_1$ or other measures  discussed in this paper.

The effect of pure loss and loss plus device imperfections  may be very detrimental \cite{ma08} {and} must be  fully quantified for both KCQ and BB84 type protocols with a proper criterion.

{Note that coherent or squeezed states of considerable energy cannot be used in BB84 type protocols to alleviate loss, due to the \emph{signal discrimination attack} Eve may launch near the transmitter. Such attack is thwarted in KCQ protocols by the shared secret key.}

\subsection{System Integration and Implementation}
It is difficult to implement BB84 type cryptosystems close to the protocol prescription due to the high performance devices required. In contrast, KCQ qumode protocols require only off-the-shelf optical technology. Furthermore, conventional amplifiers can be used on them up to a certain number \cite{yuen03,liang05} depending on the system. They can also be readily integrated with existing optical networks. All of these are difficult with the weak-signal BB84 type protocols.

\section{Conclusion}
Physical cryptography, including KCQ direct encryption as well as BB84 and KCQ key generation, employs secrecy protection mechanisms at the physical signal level away from the bit level at the application layer end of a communication link. It cannot be attacked from such end and Eve has to physically intercept the transmission link with sophisticated technology in order to launch any meaningful attack. This automatically rules out ``petty thefts'' and constitutes a  significant security advantage compared to standard techniques, similar to digital versus analog wireless rf transmissions. Apart from the possibility of rigorous security proofs, which has to be tempered by the corresponding problem of adequate physical modeling, physical cryptography offers a totally new way of securing privacy different from all the standard high-rate  cryptographic techniques in use. It is a ``new paradigm'' in cryptology.

A major implication of our KCQ approach to BB84 type approach is that a PRNG should be used to generate a running key that determines the users' choice of basis as described in Section II. This should be done even when intrusion level estimation is still employed to retain some BB84 feature for a weak signal or qubit protocol. There are many resulting advantages both from a practical implementation and a theoretical security analysis point of view.

The KCQ approach itself seems to hold great promise. Under universal heterodyne attack, we have shown that in principle fresh key generation is quite possible in the CPPM system of Section VI with respect to the attacker's total probability profile.

Finally, it is well to recall that we still need to develop a meaningful and sufficiently {strong} security measure that can be usefully estimated {and achieved} in concrete realistic protocols.

%

\appendices
\section{Inadequacy of Exponentially Small Information for Eve}

The strongest theoretical security claims (proofs) that have been offered thus far in QKD is that Eve's total mutual information on the $n$-bit generated key $G$ is exponentially small in $n$ in various BB84 type protocols. Here we will show in what ways this claim is insufficient for operationally meaningful security guarantee. The criterion of vaiational distance from a uniform string instead of mutual information is quantitatively similar and discussed in Appendix B. We are not talking here about the composition problem or issues of system modeling. It is purely the quantitative security guarantee within the system model.

Let $I_E = n -H_E(G)$ be Eve's total information on the $n$-bit generated key $G$. The well known quantitative claim is that for large enough $n$,
\begin{equation} \label{eveinfobound}
I_E \leq 2^{-\lambda n}
\end{equation}
for some function $\lambda$ of the system parameters. This $I_E$ is an average over various random parameters. If we let $r_i$ be the possible values of such parameters with probability distribution $p(r_i)$, the $I_E$ in (\ref{eveinfobound}) is of the form
\begin{equation} \label{aveinfo}
I_E = \sum_i p(r_i) I_E(r_i)
\end{equation}
where $I_E$ is Eve's information for a given $r_i$. In particular, it is averaged over Eve's observations $y_n^E$ that gives her a {specific} conditional probability distribution (CPD) $\{p_i\}, p_i = p(g_i|y_n^E)$ for $i \in \overline{1-N}, N =2^n$. Thus, for a given value of average $I_E=I_0$ in (\ref{aveinfo}) satisfying (\ref{eveinfobound}), there is at least one but generally many values of $y_n^E$ with $I_E(y_n) \equiv I(\{p_i\}) = I_E(p(g_i|y_n^E))$ that exceeds $I_0$. A reasonable guess would be that roughly half of the times $I_E(y_n)$ exceeds its average value $I_0$. For these values of $y_n$, the constraint (\ref{eveinfobound}) would not apply, i.e., $I_E(y_n) > I_0 \leq 2^{-n}$. Without an estimate of the probability on this set of $y_n$ values, the security guarantee is somewhat shaky.

This point between average versus worst case also occurs in almost all ``computational'' problems, say on the average versus worst case complexity, the latter usually taken to be the more appropriate measure. For cryptographic security, it is clear that the worst case also should be considered, as it typically was in various QKD considerations. {Using the Markov inequality (\ref{markovinequality}) to handle the randomness in $I_E$ would induce a much more stringent requirement $I_E \leq 2^{-2\lambda n}$ than (27).}

{Assume} $I_E(y_n)$ satisfies (\ref{eveinfobound}), we have seen in Section III.B that Eve's maximum probability of getting the whole $G$ correctly could be as big as $p_1 \sim 2^{-\lambda(n +\log n)}$ from (10). Also, from (13), Eve's maximum probability of getting a fraction $0 < f \leq 1$ of the $n$ bits in $G$ correctly is $p_1(f) \sim 2^{-\lambda(n +\log n)} / f$. These are adequately small if $\lambda \sim 1$ for the $n$-bit $G$. For $\lambda \ll 1$ or even $\lambda \sim 0.5$, it is \emph{not clear in what sense } $G$ is close to an $n$-bit uniformly random string. The problem is especially acute when $G$ is used to serve as one-time pad key. At best, it is a $\lambda n$-bit `nearly uniform' string and calling it an $n$-bit fresh key is an unwarranted exaggeration.

What are the possible $\lambda$'s one can obtain, in principle as well as in practice? There are few theoretical papers \cite{inamori01,hamada04,biham06,hayashi06,hayashi07} that give an explicit expression for $\lambda$, none of which gives it a sufficiently explicit form to tell readily whether $\lambda \sim 1$ can be achieved with what system parameters. The situation is a lot worse in practice. The best reported $I_E/n$ appears to be that of ref. \cite{hasegawa07}
 for $n$ up to $10^6$ bits with $I_E/n \sim 2^{-9} \sim 10^{-3}$. Many possible disastrous breaches of security are not ruled out with such numbers. In addition to the possibility of Eve's getting the whole $10^6$ bits $G$ with $p_1 \sim 10^{-3}$, from (13) some $10\%$ of $G$ or $10^5$ bits may possibly be obtained with probability $10 p \sim 10^{-2}$, or some $10^4$ bits of $G$ with probability $100p \sim 10^{-1}$, in addition to $10^3$ bits with {probability $1$}.

\section{Problem of the Variational Distance Security Measure}
In Section III and Appendix A, we showed that Eve's mutual information $I_E$ is not a good measure of security on an $n$-bit string $G$ unless $I_E/n \sim 2^{-\lambda n}$ for {a sufficiently large $\lambda$}. In addition to $I_E$, the variational distance $\delta_E$ of Eve's conditional probability distribution (CPD) from the uniform random variable $U$ with $p(u_i) = 2^{-n}, i \in \overline{1-N}, N =2^n$, {may be} used as security measure. By definition, the variation distance (or statistical distance or Kolmogorov distance) between two  distributions $P$ and $Q$ over a set $\mathcal{N}$ is
\begin{equation}
\delta(P,Q) = \frac{1}{2} \sum_{i \in \mathcal{N}} |P(i)-Q(i)|.
\end{equation}
We will show that the distance $\delta_E= \delta(G,U)$ between Eve's CPD and the uniform distribution $U$ {also has a problem}.

It was suggested that $\delta_E$ is a good measure of security because $\delta(P,Q)$ ``can be interpreted as the probability that two random experiments described by $P$ and $Q$ respectively, are different'' \cite{renner08,renner05}
, an interpretation repeated in refs. \cite{konig07,konig05}. The justification for the interpretation is given by lemma $1$ in refs \cite{renner05,konig05} which states that for any two distributions $P,Q$ for two random variables $X$ and $X'$, there exists a joint distribution $P_{XX'}$ that gives $P,Q$  as marginals with
\begin{equation}
\mathrm{Pr} [X \neq X'] = \delta(P,Q).
\end{equation}
However, to the extent it makes sense to talk about such a joint distribution, the interpretation would obtain only if ``there exists'' is replaced by ``for every''. This is because since there is \emph{no} knowledge on such joint distribution, one cannot assume the most favorable case via ``there exists'' for security guarantee {or for general interpretation.} Indeed, it is not clear at all what realistic meaning can be given or claimed for the realization of such a joint distribution, other than the independent case $P_{XX'} = P \cdot Q$. In such case, even if both $P$ and $Q$ are the same uniform distribution so that $\delta(P,Q)=0$, {we have} $\mathrm{Pr}[X \neq X']= 1 - \frac {1}{N}$ and the two sides of (30) are almost as far apart as it could be since both are between $0$ and $1$. {This provides a counter-example to the interpretation.}

As a numerical measure, $\delta_E$ { suffers  the same $p_1$ problem } as $I_E$ from the fact that
\begin{equation}
\delta_E = 2^{-l} - \frac {1}{N},
\end{equation}
when Eve's CPD has $p_1 =2^{-l}, p_2 = \cdots = \frac {1-2^{-l}} {N-1}$. See Section III.B and Appendix A. When $l$ is not close to $n$, the security risk of a $\delta_E$ guarantee may be tremendous for any $n \gg l$ exactly as in the case of the $I_E$ guarantee.

{However, the subset probability guarantee of $\delta_E$ is  better than that of $I_E$. As stated in section III.C, the incremental probability from uniform for any $m$-bit subsequence is no more than $\epsilon$ for $\delta_E \leq \epsilon$. On the other hand, for $\epsilon \gg 2^{-m}$, the generated key $G$ is still far from perfect.}

In contrast to $I_E$ which can be bounded via Holevo's inequality, there is no known way to guarantee $\delta_E \leq \epsilon$ for all possible measurements from Eve. In \cite{scarani08,konig07} a strong claim is made on a quantum quantity $d$ that if $d \leq \epsilon$ the generated key $G$ is disentangled from Eve's probe and identical to $U$ except with probability $\epsilon$. However, the claim {was incorrectly drawn on the basis of (30) in these references}.  Accordingly, the universal property of such a key {as well as the quantitative security significance of $d$ do not follow as consequences.} These issues will be discussed in detail in another paper.


\section*{Acknowledgment}

I would like to thank E.~Corndorf, W.Y.~Hwang, P.~Kumar, J.~Myers, and R.~Nair for useful discussions. This work was supported by DARPA and AFOSR.

\ifCLASSOPTIONcaptionsoff
  \newpage
\fi


\begin{IEEEbiographynophoto}{Horace P.~Yuen}
is Professor of Electrical Engineering and Computer Science and Professor of Physics and Astronomy at Northwestern University. He received his degrees in Electrical Engineering from Massachusetts Institute of Technology. His technical research interest are mainly in the areas of communication and cryptography, especially those with quantum effects. He is a recipient of the 2008 IEEE/LEOS Quantum Electronics Award, the 1996 International Quantum Communication Award, a Fellow of the American Physical Society, and a senior member of the IEEE. Several of his papers are collected in various special volumes, including ``One Hundred Years of Physical Review'' which was published by the American Physical Society in 1993.
\end{IEEEbiographynophoto}





\begin{thebibliography}{1}
\bibitem{note1} The term ``BB84 type protocols'' would be used to cover variants including Ekert protocols that employ quantum entanglement explicitly.

    From the point of view of this paper, the key feature of such protocols is that intrusion level estimation via public exchange is used to bring about security.
\bibitem{note3}
She is given only a single copy, not a ``cloning machine'' {that does not exist according to the principles of quantum mechanics}. In
single-photon BB84, one such copy is enough to completely
undermine its security.
\bibitem{note4}
Note that just non-identical qubit-by qubit probes is already classified as
a joint attack in the BB84 context.
\end{thebibliography}
\end{document}